\documentclass[aps,prl,twocolumn,showpacs]{revtex4-1}  %longbibliography
\usepackage{graphicx} % Include figure files
\usepackage{amssymb}
\usepackage{float}
\usepackage[T1]{fontenc}
\begin{document}
\title
{\bf  An emergent force in a bilayer superfluid Bose-Fermi mixture}
\author{Mehmet G\"{u}nay} \email{gunaymehmt@gmail.com}
\affiliation{Department of Nanoscience and Nanotechnology, Faculty of Arts and Science, Burdur Mehmet Akif Ersoy University, 15030 Burdur, Turkey}
%\affiliation{ Institute  of  Nuclear  Sciences, Hacettepe University, 06800 Ankara, Turkey}
\break

%\onecolumngrid
\begin{abstract}
We investigate a system of two-atomic species in mixed dimensions, in which one species is spread in a three-dimensional space and the other species is confined in two parallel layers. The presence of atoms in 3-dimensions creates an induced potential for the ones confined in layers. Depending on the effective scattering length and the layer separation, the formation of p-wave pairing within the same layer or s-wave pairing between different layers has been suggested. It is shown that these pairs cannot coexist when time-reversal symmetry (TRS) is on, and there appears a transition from p-wave to s-wave as the ratio of the layer separation and the effective scattering length decreases.
With the formation of the inter-layer pairing, we find an emergent force to be present at the critical point and show that it can be derived from the ground state energy. This result offers a tool for experimentally realizing such transitions, and can find notable potential in the field of quantum-thermodynamics.
\end{abstract}
%\pacs{xxx}
\maketitle
The mechanism of Cooper pairing beyond the conventional BCS spin singlet state remains one of the most intriguing problem in condensed matter physics. After the discoveries of new phases of $ ^3 $He~\cite{PhysRev.123.1911,PhysRevLett.28.885}, the examples of the unconventional pairings have been seen in heavy fermion~\cite{PhysRevLett.43.1892} and high temperature~\cite{bednorz1986possible} superconductors. In recent years, the mechanisms of the unconventional pairing has been a strong focus of attention in the context of superconductors having nontrivial topological properties~\cite{PhysRevB.79.094504}. These materials are important for the device applications due to having gapless edge state modes~\cite{sato2017topological}.

There are several techniques to reveal the symmetry of the order parameters experimentally. These are mostly related to the measurements of the thermodynamical quantities such as temperature dependence of the superfluid density, the specific heat, etc. In most of these observations, the existence of nodal structure in the superfluid gap play crucial role. For instance, $ T^3 $-dependence of the specific heat is considered as the existence of the point node in the superfluid gap, whereas $ T^2 $-dependence is the mark for the line nodes~\cite{bauer2012non}. Although, one can have an idea about the nature of the superfluid gap, it is difficult to derive the underlying mechanism of the pairing from these measurements.  

At this point, atomic gases provide an ideal platform for simulating exotic physical systems due to having the ability to control the system parameters externally. Very recently, experimental observations of the pairing in two-dimensional Fermi gases have been reported through the BEC-BCS crossover~\cite{PhysRevLett.106.105301,PhysRevLett.108.045302,PhysRevLett.114.230401,feld2011observation,PhysRevA.94.031606}. In addition to these two-dimensional systems, there are some promising setups in mixed-dimensions for the observation of superfluid gap with different symmetries~\cite{PhysRevA.95.053633}. The long-range interaction in obtaining unconventional pairing with such symmetries is a crucial ingredient. In this direction, it has been shown that in a system of two-atomic species in mixed dimensions, in which one species is confined in two-dimensions with keeping the other species in a three-dimensional space, the long-range correlations among the confined particles can be induced through the interaction with
the background particles in a three-dimensional space~\cite{PhysRevLett.121.253402,PhysRevA.94.063631,PhysRevA.61.053601,PhysRevA.61.053605,PhysRevA.96.033605,PhysRevA.82.011605}. 

In this paper, we study the thermodynamics of a system of the two-atomic species in mixed-dimensions, where identical fermions are confined in two parallel layers and placed in a three-dimensional space occupied by the BEC particles. In Ref.~\cite{PhysRevA.82.011605}, it is demonstrated that this structure can hold at least four distinct quantum phases. Moreover, in Ref.~\cite{PhysRevA.96.033605}, it is shown that the controllable topological phase transition can be observed by changing the interlayer distance and~(or) the BEC coherence length. In both of these studies, it is found that when the layers are far apart, the superfluid gap can be obtained with $p$-wave symmetry. On the other hand, with decreasing layer separation, interlayer pairing comes into play with $s$-wave symmetry. When TRS is manifested, the first-order phase transition in the superfluid gap~($s$-$p$ switching) at a critical layer separation is observed. Here, we focus on this solution to investigate the effects of such a phase transition in the superfluid gap on the thermodynamical quantities. We find that the change of the entropy is not continuous at the critical point. We also discuss that the similar behavior can be observed in the measurements of the specific heat and the density of state. These results can be used as a tool for the experimental detection of the symmetry of the superfluid gap and possible phase transitions.

Next, we study the internal energy of the system, in which the interlayer distance dependence of the ground state energy is clearly shown. By using this relation, we define a force emerging with the formation of the interlayer-pairing. Since the induced potential is long-range, the nature of the emerging force also appears long-range when the layer separation is smaller than the critical point~(D$_c$) and it vanishes beyond that point~(D>D$_c$). This result~(interlayer pairing force) can be realized in excitonic condensate systems~\cite{eisenstein2004bose,hakiouglu2014measurable} as well as the bilayer graphene structures~\cite{lu2019superconductors} when the sufficient pairing between different layers is present. Finally, we discuss that the work done by this force can be used in the realization of the quantum heat-engines. 

\begin{figure}
%\vskip-0.1truecm
\begin{center}
\includegraphics[width=50mm]{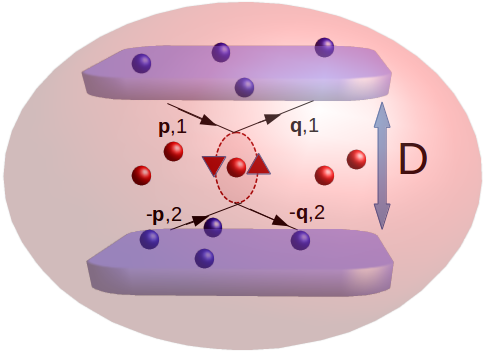}
\caption{Sketch of the system under consideration in this work. Fermions~(blue) are confined in two parallel layers with distance D, interacting through the excess of BEC~(red) backround.}
\label{fig1}
\end{center}
\end{figure}

We consider the structure as studied in Refs.~\cite{PhysRevA.96.033605,PhysRevA.82.011605}, where identical fermions are confined in two parallel layers locating at z = 0 and z = D with an equal number of particles. The layer structures are immersed in a 3-dimensional space occupied by bosons~(Fig.~\ref{fig1}) with mass $ m_B $ and denstiy $ n_B $. The BEC particles in a 3-dimensional space behaves like the interaction-center for the fermions confined in layers and make intra- and inter-layer coupling possible. The resulting the induced interaction for the fermions is given by~\cite{PhysRevLett.121.253402,PhysRevA.94.063631,PhysRevA.61.053601,PhysRevA.61.053605}
\begin{eqnarray}
{ V}_{ \rm ind}^{\nu \nu^\prime}({\bf{p}})=-\frac{2g^2n_B m_B}{\sqrt{{\bf{p}}^2+2/\xi_B^2}} e^{-{\rm D}|\nu-\nu^\prime|\sqrt{{\bf{p}}^2+2/\xi_B^2}},
\label{Eq:Potential}
\end{eqnarray}
where $ g $ represents Bose-Fermi coupling and $ \xi_B=(8 \pi n_B m_B)^{-1/2} $ is the coherence length. Since the induced interaction is long-range and attractive, it opens a gate for the fermions to form Cooper pairing both in between the layers and within the each layer. This can be observed from the form of the induced potential, which is labeled with the layer index $ \nu=1,2 $. This layer index introduces an extra degree of freedom and makes it possible to observe the superfluid gap with different symmetries~\cite{PhysRevA.96.033605,PhysRevA.82.011605}. 

With the formation of such pairs, the Hamiltonian in the basis: $\Psi_{\bf k}^\dagger=(\hat{c}_{{\bf k} 1}^\dagger~\hat{c}_{{\bf k} 2}^\dagger~\hat{c}_{-{\bf k} 1}~\hat{c}_{-{\bf k} 2})$ can be given by 
\begin{eqnarray}
{H}=\sum_{\bf k} \Psi_{\bf k}^\dagger {\cal H}_{\bf k} \Psi_{\bf k}, \qquad {\cal H}_{\bf k}=\pmatrix{\epsilon_k \sigma_0 & \Delta_{\bf k} \cr \Delta_{\bf k}^\dagger & -\epsilon_k\sigma_0}\, ,
\label{ham_1}
\end{eqnarray}
where $ \hat{c}^\dagger_{{\bf k} j} $~($ \hat{c}_{{\bf k} j} $) is the creation~(annihilation) operator for the $ j $th layer with $ j $=1,2. Here $\epsilon_k=\hbar^2 k^2/2m-\mu$, $m$ is the band mass, $\mu$ is the chemical potential. Throughout this paper, we assume that each layer has the equal number of particles and we ignore the effects of the fermions on the BEC particles~\cite{PhysRevLett.121.253402,PhysRevA.94.063631,PhysRevA.61.053601,PhysRevA.61.053605}. The layer-indexed full $2\times 2$ matrix $\Delta_{\bf k}$ is  given by 
\begin{eqnarray}
\Delta_{\bf k}=\pmatrix{\Delta_{1 1}({\bf k}) & \Delta_{12}({\bf k}) \cr \Delta_{21}({\bf k}) & \Delta_{22}({\bf k})\cr},
\label{Delta}
\end{eqnarray} 
where we obtain the self consistent components with the mean field approach as   
\begin{eqnarray}
 \Delta_{\nu \nu^\prime}({\bf k})=-\frac{1}{\cal V}\,\sum_{\bf q}{ V}_{ \rm ind}^{\nu \nu^\prime}({\bf k-q})\,\langle \hat{c}_{{\bf q},\nu}\,\hat{c}_{{\bf -q},\nu^\prime}\rangle .
 \label{OPs}
 \end{eqnarray}
Here, we focus on the solutions of the gap equation when TRS is manifested. Moreover, the intralayer~(triplet) pairs are connected with this symmetry as: $\Delta_{11}({\bf k})=\Delta^*_{22}(-{\bf k})$ and we define: $\Delta_{11}({\bf k})=\Delta_t(k) e^{i(\phi_{_{\bf k}}-\phi_{_0})}\propto (k_x+ik_y)$, where $ \Delta_t(k) $ is a real and even function with $ k= |{\bf k}|$ and $ \phi_{_0} $ is the phase difference between the particles residing in the upper and lower layers. We take the interlayer pairing to be $s$-wave so that
$\Delta_{1 2}({\bf k})=-\Delta_{21}({\bf k})=\Delta_s(k)$~\cite{PhysRevA.96.033605}. Here, $\Delta_s(k)$ is a even function due to the Fermi antisymmetry and the presence of TRS dictates that it can only take real values. After some algebra, we obtain the corresponding self-consistent equations for the triplet and the singlet amplitudes as
\begin{small}
\begin{eqnarray}
\Delta_t(k)&=&-\frac{1}{\cal V}\,\sum_{k^\prime,\lambda}{\rm V}_t(k, k^\prime)\, \frac{\tilde{\Delta}_{k^{\prime}\lambda}}{4E_{k^{\prime}\lambda}} {\rm tanh}\bigg(\frac{E_{ k^\prime \lambda}}{2 k_B T}\bigg), \\ ~~~~
\label{Eq:triplet}
\Delta_s(k)&=&-\frac{1}{\cal V}\,\sum_{ k^\prime,\lambda}{\rm V}_s(k, k^\prime)\, \frac{\lambda \tilde{\Delta}_{k^{\prime}\lambda}}{4E_{k^{\prime}\lambda}} {\rm tanh}\bigg(\frac{E_{ k^\prime \lambda}}{2 k_B T}\bigg),~~~~
\label{Eq:singlet}
\end{eqnarray}
\end{small}
and for the total number 
\begin{small}
\begin{eqnarray}
N=\frac{1}{2}\sum_{\bf k \lambda} \bigg[1-\frac{\epsilon_k}{2 E_{\bf k \lambda}} {\rm tanh}\bigg(\frac{E_{\bf k\lambda}}{2 k_B T}\bigg)\bigg].
\label{number_1}
\end{eqnarray}
\end{small}Here $\lambda=\pm$ is the branch index, $ {\rm V}_t(k, k^\prime)=\langle{{ V}_{ \rm ind}^{\nu \nu}}({\bf |k-k^\prime|}) \cos{({\tilde{\phi}})}\rangle_{\tilde{\phi}}$  represents the corresponding interaction channel for the triplet and  $ {\rm V}_s(k, k^\prime)=\langle{{ V}_{ \rm ind}^{1 2}}({\bf |k-k^\prime|})\rangle_{\tilde{\phi}}$ for the singlet pairing amplitudes respectively with the angular average of the relative phase $\tilde{\phi}=\phi_{_{\bf k^\prime}}-\phi_{_{\bf k}}$. Here, one can observe that the short-range interaction~(i.e. $ { V}({\bf |k-k^\prime|})=V_0$ ) is insufficient to create pairing in the triplet channel due to the appearence of $\cos{({\tilde{\phi}})}$ term. And, the existence of such terms, in addition to induced potential, will be eliminated through the angular average. The eigen-energies with the new pairing fields, $ \tilde{\Delta}_{k\lambda}=\Delta_t(k) + \lambda  \Delta_s(k)$, are given by
\begin{eqnarray}
E_{k\lambda }=\sqrt{\epsilon_k^2 + \tilde{\Delta}^2_{k\lambda}}.
\label{spectrum} 
\end{eqnarray}

\begin{figure}
%\vskip-0.1truecm
\includegraphics[scale=0.65]{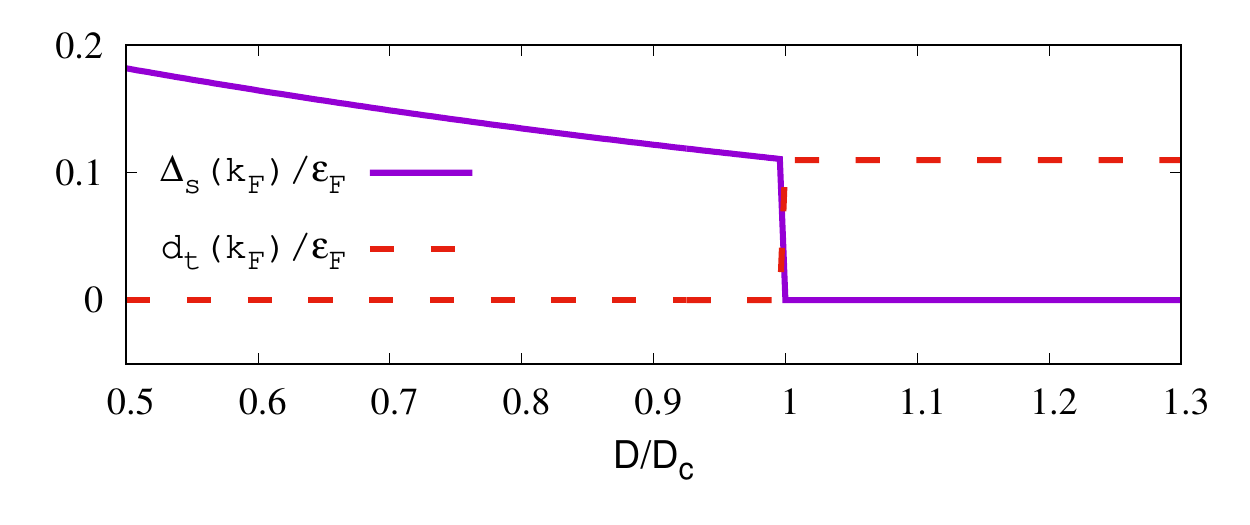}
\vskip-0.2truecm
\caption{The amplitudes of the singlet~$  \Delta_s(k_f) $ and the triplet~d$  _t(k_f)=\Delta_t(k_f)/k_f $ pairs with respect to the layer separation D. $s$-$p$ switching can be observed at D = D$_c$. }
\label{fig2}
\end{figure}

The ratio of the layer separation and the coherence length~($d$ = D/$ \xi_B $) has decisive influence on the symmetry of the superfluid gap. When this ratio is large D$ \rightarrow \infty$, the induced interaction for the singlet pairing is suppressed. Therefore, in this limit, triplet solutions are favored and the superfluid gap is expected to have $p$-wave symmetry~\cite{nishida2009induced}.  On the other hand, when D$ \rightarrow 0$, it was shown that the symmetry of the gap can only be $s$-wave if the TRS is manifested~\cite{PhysRevA.96.033605,PhysRevA.82.011605}. In Fig.~\ref{fig2}, we demonstrate this phase transition~($s$-$p$ crossing) as a function of the layer separation, D. The similar result is also obtained in Ref.~\cite{PhysRevA.96.033605}~(see Fig. 3(b) therein), where they investigate the topological phase transtion of such crossing. Since the phase transition is of first-order, we rewrite the new pairing field as 
\begin{small}
\begin{eqnarray}
\tilde{\Delta}_{k\lambda}\equiv\tilde{\Delta}_{k} ={\Delta}_s(k,{\rm D}) \Theta({\rm D}_c-{\rm D})+{\Delta}_t(k) \Theta({\rm D}-{\rm D}_c),
\label{step} 
\end{eqnarray}
\end{small}where \begin{small}$ \Theta({\rm D}-{\rm D}_c) $\end{small} is the unit step-function. And, tuning the distance between layers around D$_c$ can lead to dramatic changes in the thermodynamic quantities of the system due to altering symmetry of the superfluid gap. This can be done in an experiment by adiabatic changing the scattering length $ \xi_B $~\cite{PhysRevLett.101.040402,PhysRevA.82.033609} around $ d_c $ = D$_c/\xi_B$ via Feshback resonances, which is more feasible than the moving layers up and down. 

\begin{figure}
%\vskip-0.1truecm
\includegraphics[scale=0.6]{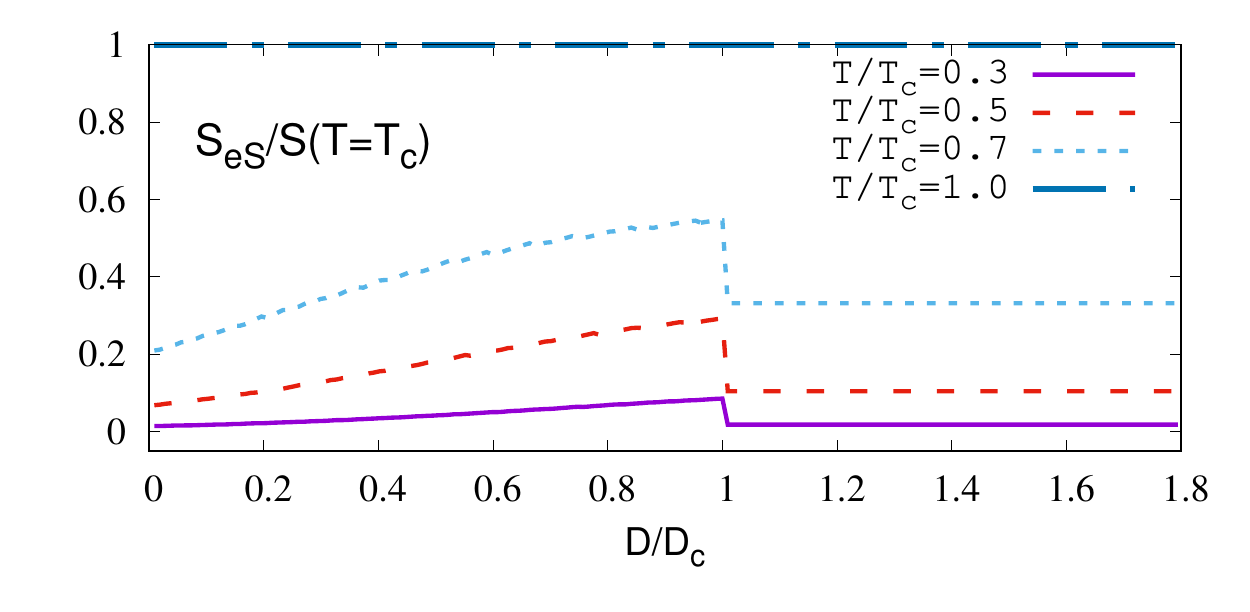}
\caption{The behaviour of the entropy with respect to the layer separation for various temperature. The jump can be observed at the critical layer separation  D$_c $ for  T$<$T$_c $. We scale the entropy with its value at the critical temperature S(T=T$_c$).}
\label{fig3}
\end{figure}

In the following, we show that the appearence of the inter-layer pairing costs to the thermodynamical quantities. For instance, in Fig.~\ref{fig3}, we demonstrate the reflection of the $s$-$p$ crossing in the entropy of the system, where  we calculate it from~\cite{tinkham2004introduction} 
%and suggest a route for the experimental observation of such phase transition. 
\begin{eqnarray}
S=-2 k_{B} \sum_{\bf k} [(1-f_k) {\rm ln}(1-f_k)+f_k {\rm ln}(f_k)].
\label{entropy} 
\end{eqnarray}
Here $ f_k=(1+e^{\beta E_k})^{-1} $ is the Fermi-Dirac factor with $ \beta=1/k_BT $. Since the pairing fields cannot coexist due to TRS, the energy branches become degenerate, i.e., $ E_{k\lambda }=E_{k} $. It can be seen from Fig.~\ref{fig3} that there appears a jump in the entropy of the system at critical layer separation D$ _c $, which supports our theory. The similar jump can also be observed in the related thermodynamic measurements. For instance, by using the relation: $C_v=T dS/dT$, it is natural to expect similar behavior in the specific heat.  Additionally, the altering symmetry of the superfluid gap can be readily detected through the density of state measurement, in which it vanishes continuously as the energy goes to zero~($E\rightarrow 0$) for the $p$-wave gap, whereas, in a region~$E<\Delta_s(k_f)$, no state is available for the isotropic $s$-wave gap.

\begin{figure}
\vskip-0.7truecm
\includegraphics[scale=0.6]{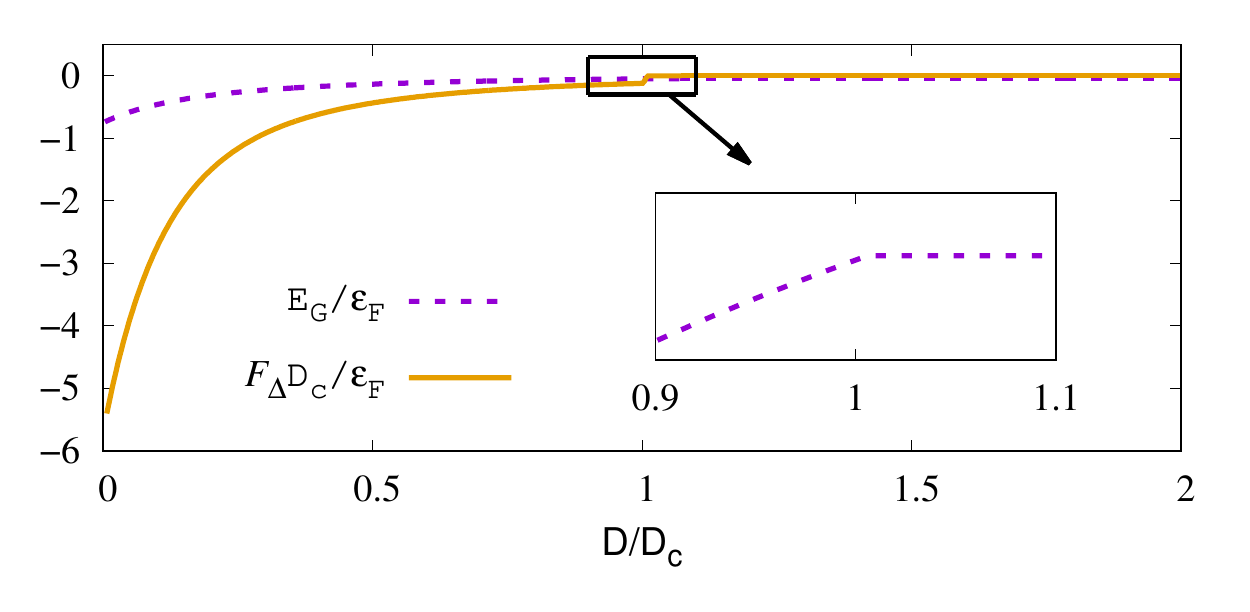}
\caption{ The behavior of the ground state energy~(dotted-line) and the inter-layer pairing force~(straight-line) with respect to the layer separation at T = 0. The long-range nature of the force can be observed with the formation of the interlayer pairing for D<D$_c$ and it vanishes when the layer separation exceeds the critical point. The inset magnifies the ground state energy around D$_c$.}
\label{fig4}
\end{figure}

Next, we investigate the ground state energy of the system, which is the key function of interest in this paper, defined at zero temperature by 
\begin{eqnarray}
E_{G}=\sum_{\bf k} \bigg( \epsilon_{{\bf k}}-E_{{\bf k}}-\frac{\tilde{\Delta}_{\bf k}^2}{2E_{{\bf k}}} \bigg),
\label{potential} 
\end{eqnarray}
where the last term comes from the mean-field solution. It is apparent from Eq.~(\ref{potential}) that the ground-state energy is dependent on the layer separation through the superfluid gap. An essential consequence of the relation between the energy and the distance is the force. Therefore, one can speculate appearing an emergent force engaged with these variations in the internal energy of the system, as such observations are related to the distance between the layers. 
This type of force, which emerges with the formation of inter-layer pairing, can be derived by taking the derrivative of the ground state energy with respect to layer separation. It is defined by
\begin{eqnarray}
{\textit{F}_\Delta}=-\frac{\partial E_{G}}{\partial {\rm D}}.
\label{force} 
\end{eqnarray}
In Fig.~\ref{fig4}, we demonstrate the results of the Eq.~(\ref{potential}) and Eq.~(\ref{force}) by varying layer separation. The phase transition in the superfluid gap can be obtained by minimizing the free-energy of the system. It can be read from the inset of the Fig.~\ref{fig4} that the lower-energy is present when the layer separation is smaller than the critical value, which makes the interlayer pairing more favourable. Moreover, the nature of the force~(straight line), in a region D$ < $D$_c$, is long-range and  decays with increasing D and vanishes when the layer separation exceeds the critical value. This result is expected from the form of the induced interaction in Eq.~(\ref{Eq:Potential}). 

In obtaining results, we use similar parameters with Ref.~\cite{PhysRevA.96.033605} and scale momenta with Fermi momentum~$ k_F=\sqrt{4 \pi n_F} $, where $ n_F $ is the fermion density in each layer. We consider a weak Bose-Fermi coupling $ g=2\pi a/\sqrt{m_r m_{_B}} $, where  $ a k_F=0.12 $ is the scaled-scattering length~\cite{PhysRevLett.101.170401}, which is a tunable parameter via Feshbach resonance and $ m_r $ is the reduced mass.

The presence of the symmetries considered in this work lead to have a first-order phase transition in the superfluid gap, which enhances the results of this paper. However, in a system where the second-order phase transition is the case, the effects of the layer separation will also be present, as there always be a critical D$ _c $~(see Fig. 2(a) in Ref.~\cite{PhysRevA.96.033605}). For instance, in the case of TRS breaking, the transition in the superfluid gap from $p$-wave to mixed parity symmetry can be observed at the critical layer separation. In analogy with the relation, e.g., between the temperature and the specific heat, where a jump can be observed in specific heat at the critical temperature, the formation of the inter-layer pairing will leave a mark in the thermodynamic quantities. And, by following these signs, the findings can help to identify the nature of the superfluid gap. 

Finally, let us briefly discuss how the results obtained in this paper can be implemented to the statistic of the work done on a quantum system when there is a phase transition in the superfluid gap. Assume that the system is initially prepared with the ratio $d_0~(>d_c)$ and moved to the final value $d_1~(<d_c)$. The work done on the system can then be defined as 
\begin{eqnarray}
W=E_{G}(d_1)-E_{G}(d_0).
\label{work} 
\end{eqnarray}
If such process exhibits the quench protocol, the work $W$, is rather characterized by a probability distribution $P(W)$~\cite{PhysRevLett.78.2690,PhysRevE.75.050102}, in which the characteristic function can be given by~\cite{PhysRevLett.101.120603}
\begin{eqnarray}
{G}(t)=\int dW e^{iWt} P(W),
\label{LS_echo} 
\end{eqnarray}
where its connection with the Loschmidt echo was shown in Ref.~\cite{PhysRevLett.101.120603}, as $ {G}(t) =[{\cal G}(t)]^\ast$. Here, the amplitude is given by ${\cal G}(t)=\langle e^{iH(d_0)t} e^{-iH(d_1)t} \rangle  $, and the avarage can be taken by using the initial equilibrium density matrix, $\rho_0=exp[-\beta H(d_0)]/Z$ with Z being the partition function. If one defines the eigenstates of $H(g_0)~\{H(g_1)\} $ as $|\psi_n\rangle~\{|\phi_m\rangle\}$, the probability distribution can be ontained as~\cite{PhysRevLett.101.120603}
\begin{eqnarray}
 P(W)=\sum_{n,m} \delta(W-(E_m-E_n)) |\langle \psi_n|\phi_m\rangle|^2 P_m,
\label{Dist} 
\end{eqnarray}
where $P_m={\rm exp}(-\beta E_m)/Z$. With further efforts, the bilayer superfluid structures can serve to test the work fluctuation theorems as shown above. Moreover, the efficieny of the work done by the interlayer pairing force can also be tested in Otto cycles in realization of the quantum heat engine applications.

In summary, we study the bilayer superfluid Bose-Fermi mixture in a mixed dimension and show that it is possible to reveal the nature of the superfluid gap by following the thermodynamical signatures. Moreover, it is found that the formation of the inter-layer pairing creates an additional force. This force, actually, will be present for any bilayer system as long as the sufficient pairing between different layers is observed. We also discussed that, besides the fundamental interest, the work done by this force can be used in the quantum-heat engine applications. Such solutions in these structures are, in general, considered in terms of the Casimir force or pressure~\cite{PhysRevLett.83.1187, PhysRevLett.111.055701}. The addition of the interlayer pairing force can enrich the problem and lead to explore more exotic stuctures with variety of the applications.

%\section*{Acknowledgments}
%The author thanks to Georg Bruun and \"{O}zg\"{u}r Esat M\"{u}stecapl\i o\u{g}lu for useful  comments.
%This research was supported by  The Scientific and Technological Research Council of Turkey (TUBITAK) Grant No. 117F118.
\bibliography{bibliography}

\end{document}